\documentclass[12pt]{article}
\usepackage{latexsym}
\usepackage{jheppub}
\usepackage{amssymb,amsfonts}
\usepackage{amsmath,amsthm,epsfig,euscript,array,cancel}
\usepackage{bbm}
\usepackage{ifpdf}
\ifpdf \hypersetup{%
  pdftitle    = {Notophs of N=8 supergravity}
  pdfkeywords = {Supersymmetry, supergravity, superspace},
  pdfauthor   = {Igor A. Bandos and Tomas Ortin},
  plainpages  = true,
  colorlinks  = true,
  citecolor   = blue,
  urlcolor    = red,
  linkcolor   = black
}
\newcommand{\hepth}[1]{{\tt
\href{http://www.arXiv.org/abs/hep-th/#1}{hep-th/#1}}}

\newcommand{\arxiv}[1]{{\tt
\href{http://www.arXiv.org/abs/#1}{#1}}} \else
  \newcommand{\hepth}[1]{{\tt hep-th/#1}}

  \newcommand{\arxiv}[1]{{\tt arXiv:#1}}
\fi 

\preprint{IFT-UAM/CSIC-15-012 \\ 
Phys.Rev. D91 (2015) 085031  }

\title{Notophs of $\mathcal{N}=8$ supergravity}

\author{
Igor Bandos $^{\dagger\ddagger}$ and Tom\'as Ort\'{\i}n$^{\ast}$
\\
$^{\dagger}$ {\it Department of Theoretical Physics, University of the Basque Country UPV/EHU, P.O. Box 644,
48080 Bilbao, Spain} \\  $^{\ddagger}$ {\it IKERBASQUE, Basque Foundation for Science, 48011, Bilbao, Spain} \\
$^{\ast}$ {\it Instituto de F\'{\i}sica Te\'orica UAM/CSIC, Calle Nicol\'as Cabrera, 13-15, C.U. Cantoblanco,
E-28049 Madrid, Spain} }

\date{February 2nd,  2015}

\abstract{ %
  We study the tensor gauge fields (``notophs'') of ungauged
  $\mathcal{N}=8,D=4$ supergravity in superspace. These are described by
  2-form potentials $B_{2}^{G}$ in the adjoint representation of
  $G=E_{7(+7)}$. The consistency of the natural candidates for the superspace
  constraints for their field strengths $H_{3}^{G}$ fixes the form of the
  generalized Bianchi identities $DH_{3}^{G}= \ldots$ and also requires the
  potentials $B_{2}^{G/H}$ with indices of $G/H=E_{7(+7)}/SU(8)$ generators to
  be dual to the scalars of the $\mathcal{N}=8,D=4$ supergravity multiplet. In
  contrast, the field strengths of the 2-form potentials corresponding to the
  $SU(8)$ generators are dual to fermionic bilinears, so that these potentials
  are auxiliary rather than physical fields. Their presence, however, is
  essential to formulate a tensor hierarchy of $\mathcal{N}=8,D=4$
  supergravity consistent with its U-duality group $E_{7(+7)}$.

}


\begin{document}

\maketitle

\setcounter{page}2

\section{Introduction}

The action of the maximal $\mathcal{N}=8,D=4$ supergravity was obtained in \cite{Cremmer:1979up} by dimensional
reduction of the $D=11$ supergravity \cite{Cremmer:1978km} followed by dualization of 7 antisymmetric tensor
gauge potentials $B_{\mu\nu}^{I}$ originating in the 11-dimensional 3-form, called ``notophs'' in
\cite{Ogievetsky:1967ij},\footnote{``Notoph'' is ``photon''
  read from the right to the left.  Other, more popular names are Kalb-Ramond
  field \cite{Kalb:1974yc}, 2-form, and even ``B-field''
  \cite{Seiberg:1999vs}.} to scalars. Then the complete set of 70(=28+35+7)
scalars of the $\mathcal{N}=8,D=4$ supergravity multiplet was found to parametrize the coset space
$E_{7(+7)}/SU(8)$ \cite{Cremmer:1979up}.

The natural question is whether this duality can be performed in an opposite direction, introducing a dual
``notoph'' for each scalar of the theory. In this paper we study this problem in the $\mathcal{N}=8,D=4$
superspace formulation of supergravity. To be more precise, we search for a ``duality symmetric'' formulation
of the theory, containing both the scalar fields and the notophs rather than trying to replace everywhere the
former by the latter (which is not possible beyond the linear approximation in fields).

The motivation for such a study is two-fold. On one hand, we hope that our results will contribute to a deeper
understanding of the U-duality group of the $\mathcal{N}=8,D=4$ supergravity, the exceptional Lie group
$E_{7(+7)}$. The interest in this symmetry has remained high during the almost 36 years passed since its
discovery in \cite{Cremmer:1979up}, and, recently, a relation with the exceptional convergence properties of
its loop amplitudes has been proposed (see Refs.~\cite{Brodel:2009hu} and references therein.)

On the other hand, the knowledge on existence of $(p+1)$-form gauge potentials in a supergravity superspace
might indicate the existence of supersymmetric extended objects, p-branes, coupled electrically to these
potentials. In this sense our results imply the possible existence of a family of supersymmetric strings in an
$\mathcal{N}=8,D=4$ supergravity superspace\footnote{The BPS
  branes of the maximal supergravity theories were studied originally in
  Refs.~\cite{Ferrara:1997ci}, but their worldvolume
  actions are, in general, unknown.}. The search for possible worldvolume
actions of such hypothetical superstrings is one of the natural applications of our results.

A first result showed by our study is that, to be consistent, one has to introduce a 2-form potential for each
of the generators of $G=E_{7(+7)}$ group, $B_{2}^{G}= (B_{2}^{G/H}, B_{2}^{H})$, and not just for the
generators of the coset $G/H$. This result can be generalized to other theories with scalars parametrizing a
symmetric space \cite{IB+TO=Next}. An early example of how the dualization of scalars requires the introduction
of a $(d-2)$-form potential for each generator of the isometry group, even though their numbers do not match,
is the dualization of the dilaton and RR 0-form of $\mathcal{N}=2B,D=10$ supergravity in \cite{Meessen:1998qm} (see also \cite{Dall'Agata:1998va}):
the two real scalars parametrize an $SL(2,\mathbb{R})/SO(2)$ coset space and they are dualized into a triplet
of 8-forms transforming in the adjoint. The existence of this triplet of 8-forms is required by the symmetry
algebra $E_{11}$ \cite{Bergshoeff:2006qw} and has clear implications in the classification of the possible
7-branes of the theory \cite{Meessen:1998qm,Bergshoeff:2006gs,Bergshoeff:2007ma,Bergshoeff:2007aa}. In the context of the embedding tensor formalism for 4-dimensional gauged supergravities
\cite{Nicolai:2000sc,deWit:2005ub,deWit:2008ta,deWit:2008gc} (bosonic, spacetime) 2-form potentials in the
adjoint representation of the duality group have to be introduced for different technical reasons, unrelated
to the dualization of scalar fields, and for the specific case of $\mathcal{N}=8,D=4$ supergravity this was
done some years ago in Refs.~\cite{deWit:2005ub,de Wit:2007mt}. The general duality rule between scalars and
$(d-2)$-forms was established in Refs.~\cite{deWit:2008gc,Bergshoeff:2009ph,Hartong:2009vc} using the embedding
tensor formalism, but the results remain valid in the ungauged limit.

The study of supersymmetrization of these and other higher-rank gauge potentials has received much less
attention\footnote{Some partial results on
  the supersymmetrization of the 2-forms dual to scalars in 4-dimensional
  $\mathcal{N}=2,1$ theories can be found in
  \cite{Bergshoeff:2007ij,Hartong:2009az}. Supersymmetry has, nevertheless,
  been one of the main tools to find higher-rank potentials that can be added
  to the 10-dimensional maximal supergravities
  \cite{Bergshoeff:2006qw,Bergshoeff:2007ma,Bergshoeff:2010mv}, in particular for $(d-1)$- and
  $d$-form potentials.} and in this paper we are going to start filling this
gap for the case of the notophs of $\mathcal{N}=8,D=4$ supergravity using the superspace formalism.  The
knowledge of the gauge and supersymmetry transformations of these fields is a key ingredient in the
construction of $\kappa$-symmetric worldvolume actions for possible associated supersymmetric string (p-brane) models.

In superspace formalism the problem of duality symmetric formulation, including the scalars and 2-form
potentials dual to them on the mass shell, can be posed as searching for a set of constraints for 3-forms
$H_{3}^{G}=dB_{2}^{G}+\ldots$ which are generalized field strengths of the corresponding 2-form potentials
$B_{2}^{G}$ defined on superspace.  Below we present such superspace constraints for the $E_{7(+7)}$ algebra
valued $H_{3}^{G}=H_{3}^{^{E_{7(+7)}}}$ on the curved $\mathcal{N}=8$ superspace of maximal $D=4$ supergravity,
study their self-consistency conditions: the generalized Bianchi identities (gBIs) $dH_{3}^{G}=\ldots$.

The explicit form of these gBIs are part of the definition of the tensor hierarchy of the Cremmer--Julia (CJ)   $\mathcal{N}=8$
supergravity.\footnote{The tensor
  hierarchy arises naturally in the democratic gauging of theories using the
  embedding-tensor formalism
  \cite{Nicolai:2000sc,deWit:2005ub,deWit:2008ta,deWit:2008gc},
  but the fields still make sense when the embedding tensor and any other
  deformation parameters are switched off, in the ungauged, undeformed
  theory.} They reflect the group theoretical structure associated to the
$E_{7(+7)}$ symmetry of $\mathcal{N}=8$ supergravity in the dual language. We are going to recover this piece
of the tensor hierarchy starting from the natural candidate for superspace constraints for $H_{3}^{G}$ and
requiring that the algebraic part of the suitable gBIs, concentrated in their lower dimensional components,
should be satisfied identically when the candidate constraints are taken into account. At this stage we find,
in particular, that the standard Bianchi identities $dH_{3}^{G}=0$, if imposed, would lead to inconsistency and
also that one cannot formulate a consistent set of constraints for the 3-forms corresponding to the coset
generators, $H_{3}^{G/H}$, without introducing simultaneously the 3-forms $H_{3}^{H}$ corresponding to the
generators of the stability subgroup $H=SU(8)$ of the coset. In this sense one of the message of this paper is
that the superspace approach can be used in search for a consistent tensorial hierarchies of supergravity (as
well as also of the theories invariant under rigid supersymmetry).

After this is done, we further study their higher dimensional components and show that the duality relations
between the field strengths of the notophs, $H_{\mu\nu\rho}^{G/H}$, and of the scalar fields of $\mathcal{N}=8$
supergravity (generalized Cartan forms $P_{\mu}^{G/H}$) are the consequences of our superspace constraints. The
field strengths of the stability subgroup generators, $H_{\mu\nu\rho}^{H}$, are found to be dual to fermionic
bilinear; this reflects the auxiliary character of the corresponding notophs $B_{\mu\nu}^{H}$.

\section{$\mathcal{N}=8$ supergravity superspace }
\subsection{Geometry of $\mathcal{N}=8$ superspace and Cartan forms of
  $E_{7(+7)}$}

Let us denote the bosonic and fermionic supervielbein forms of ${\cal
  N}=8,D=4$ superspace $\Sigma^{(4|32)}$ by
\begin{equation}
\label{EA=EaEua} E^{A} \equiv ( E^{a}, E^{\underline{{\alpha}}}) = (E^{a}, E^{{\alpha}}_i,
\bar{E}{}^{{\dot\alpha}i}) = dZ^{M} E_{M}^{a}(Z)\; .
\end{equation}
Here $Z^{M}=(x^{\mu}, \theta^{\underline{\check{\alpha}}})$ are local bosonic and fermionic coordinates of
$\Sigma^{(4|32)}$, $a=0,1,2,3$ is Lorentz group vector index, ${\alpha}=1,2$ and ${\dot\alpha}=1,2$ are Weyl
spinor indices of different chirality (see Appendix A), $i=1,\ldots,8$ is the index of the fundamental
representation of the $SU(8)$ R-symmetry group, and $\underline{{\alpha}}$ is the 32-valued cumulative index of
$SL(2,\mathbb{C})\otimes SU(8)$. In the case of world indices, only the counterpart of this cumulative index
seems to make sense (until the Wess--Zumino gauge is fixed); it is carried by the fermionic (Grassmann-odd)
coordinate $\theta^{\underline{\check{\alpha}}}$. Finally, $\mu=0,1,2,3$ is the world vector index carried by
bosonic (Grassmann-even) coordinate $x^{\mu}$.

The curved superspace of $\mathcal{N}=8, D=4$ supergravity is endowed with a spin connection $\omega^{ab}=
-\omega^{ba}= dZ^{M} \omega_{M}^{ab}(Z)$ and with the composite connection of the $SU(8)$ R-symmetry group,
$\Omega_{i}{}^{j}= - (\Omega_{j}{}^{i})^*= dZ^{M} \Omega_{M\, i}{}^{j}(Z)$, $\; \Omega_{i}{}^{i}=0$; these are
used to define the $SL(2,\mathbb{C})\otimes SU(8)$ covariant derivative $D$. The exterior covariant derivatives
of the supervielbein forms are called bosonic and fermionic torsion 2-forms,
\begin{eqnarray}
\label{4WTa=def} T^{a} & := & DE^{a} = dE^{a} - E^{b} \wedge w_{b} {}^{a} = \tfrac{1}{2} E^{B} \wedge E^C
T_{CB}{}^{a} \; ,
\\
\label{4WTal=def} T^{\alpha}_{i} & := & DE_{i}^{\alpha} = dE_{i}^{\alpha} -E_{i}^{\beta} \wedge w_\beta
{}^\alpha - \Omega_{i}{}^{j}\wedge  E_{j}^{\alpha} = \tfrac{1}{2} E^{B} \wedge E^C T_{{CB\, i}}{}^{\!\!\alpha}
\; ,
\\
\label{4WTdA=def} T^{\dot{\alpha}i} & := & D\bar{E}^{\dot{\alpha}i} = d\bar{E}^{\dot{\alpha}i}
-\bar{E}^{\dot{\beta}i} \wedge w_{\dot{\beta}}{}^{\dot{\alpha}} -\bar{E}^{\dot{\alpha}j}\wedge
\Omega_{j}{}^{i} = \tfrac{1}{2} E^{B} \wedge E^{C} T_{{CB}}{}^{\dot{\alpha}i} \; .
\end{eqnarray}
Here $\wedge$ denotes the exterior product of differential forms with the basic properties $$E^{a}\wedge E^{b}=
-E^{b}\wedge E^{a}\, , \quad E^{\underline{{\alpha}}}\wedge E^{\underline{{\beta}}}=
E^{\underline{{\beta}}}\wedge E^{\underline{{\alpha}}}\, , \quad E^{a}\wedge E^{\underline{{\alpha}}}=-
E^{\underline{{\alpha}}}\wedge E^{a}\, , $$ and $d$ is exterior derivative which acts from the right (see
Appendix~\ref{app-moreonforms}).

By construction, the torsion 2-forms obey the Bianchi identities
\begin{eqnarray}
\label{TBIa} I_{3}^{a} & := & DT^{a} +E^{b}\wedge R_{b}{}^{a} = 0 \; ,
\\
\label{TBIal} I_{3}{}^{\alpha}_{i} & := & DT^{\alpha}_{i} +E_{i}^{\beta} \wedge R_\beta {}^\alpha -
R_{i}{}^{j}\wedge  E_{j}^{\alpha} = 0\; ,
\\
\label{TBIdA} I_{3}^{\dot{\alpha}i} & := & DT^{\dot{\alpha}i} +\bar{E}^{\dot{\beta}i} \wedge
R_{\dot{\beta}}{}^{\dot{\alpha}} + \bar{E}^{\dot{\alpha}j}\wedge R_{j}{}^{i} = 0 \; ,
\end{eqnarray}
which involve the curvature of the spin connection ($\omega_{\alpha}{}^{\beta}=
\tfrac{1}{4}\omega^{ab}\sigma_{ab\,
  \alpha}{}^{\beta}= (\omega_{\dot{\alpha}}{}^{\dot{\beta}})^{*}$),
\begin{eqnarray}
\label{R:=dom-omom} R^{ab} & = & (d\omega -\omega\wedge \omega)^{ab} = -R^{ba} = \tfrac{1}{2} E^{C} \wedge
E^{D} R_{DC}{}^{ab} \; ,
\\
\label{Ralbe:=} R_{\alpha}{}^{\beta} & = & \tfrac{1}{4}R^{ab}\sigma_{ab\, \alpha}{}^{\beta} = (d\omega
-\omega\wedge \omega)_{\alpha}{}^{\beta} = \tfrac{1}{2} E^{B}\wedge E^{A} R_{AB}{}_{\alpha}{}^{\beta}\; ,
\\
\label{bRff=N8} R_{\dot{\alpha}}{}^{\dot{\beta}} & = & (R_{\alpha}{}^{\beta})^{*} =
-\tfrac{1}{4}R^{ab}\tilde{\sigma}{}_{ab}{}^{\dot{\beta}}{}_{\dot{\alpha}} = (d\omega -\omega\wedge
\omega)_{\dot{\alpha}}{}^{\dot{\beta}} = \tfrac{1}{2} E^{B}\wedge E^{A}
R_{AB}{}_{\dot{\alpha}}{}^{\dot{\beta}}\; ,
\end{eqnarray}
and also the curvature of the induced $SU(8)$ connection, $R_{i}{}^{j} := d\Omega_{i}{}^{j} - \Omega_{i}{}^{k}
\wedge \Omega_{k}{}^{j} $. The compositeness of $\Omega_{i}{}^{j}$ is reflected by the fact that its curvature
is expressed as \cite{Brink:1979nt}
\begin{equation}
\label{Rij=PP} R_{i}{}^{j} = -(R_{j}{}^{i})^{*} = \tfrac{1}{3} \mathbb{P}_{iklp}\wedge
\bar{\mathbb{P}}{}^{jklp} , \qquad
\end{equation}
where $\mathbb{P}_{ijkl}$ is the covariant Cartan form of the $E_{7(+7)}/SU(8)$ coset and
$\bar{\mathbb{P}}{}^{ijkl}$ is its complex conjugate, which is also its $SU(8)$ dual up to an arbitrary
constant phase $\beta$:
 \begin{equation}
\label{bP=*P=eP} \bar{\mathbb{P}}{}^{ijkl} = (\mathbb{P}_{ijkl})^{*} = \tfrac{1}{4!}e^{-i\beta}
\varepsilon^{ijklpqrs}  \mathbb{P}_{pqrs}\; .
 \qquad
\end{equation}
The Cartan forms are covariantly closed,
\begin{eqnarray}
\label{DP=0} D \mathbb{P}_{ijkl} := d\mathbb{P}_{ijkl}- 4\Omega_{[i|}{}^p\wedge  \mathbb{P}_{p|jkl]}=0 \; ,
\qquad D \bar{\mathbb{P}}{}^{ijkl}=0 \; .
\end{eqnarray}
Some further properties obeyed by these forms can be found in Appendix~\ref{sec-onE77}.

Eqs.~(\ref{DP=0}), and (\ref{Rij=PP}) with $\mathbb{P}_{ijkl}$ obeying (\ref{bP=*P=eP})
are structure equations of the $E_{7(+7)}$ Lie group. These can be solved providing the expressions for the
covariant Cartan forms $\mathbb{P}_{ijkl}$ and $SU(8)$ connection $\Omega_{i}{}^{j}$ in terms of scalar
superfields of the $\mathcal{N}=8$ supergravity, the explicit form of which is not needed for our discussion
below.

\subsection{$\mathcal{N}=8,D=4$ superspace constraints and their consequences}

The constraints of $\mathcal{N}=8,D=4$ supergravity \cite{Brink:1979nt,Howe:1981gz} can be collected in the
following expressions for the bosonic and fermionic torsion 2-forms
\begin{eqnarray}
\label{Ta=N8} T^{a} & = &
 -iE^{\alpha}_{i}\wedge \bar{E}{}^{\dot{\beta}i}\sigma^{a}_{\alpha\dot{\beta}}\; ,
\\
\label{Tfa=N8} T^{\alpha}_{i} & = & \tfrac{1}{2} \bar{E}{}^{\dot{\beta}j}\wedge \bar{E}{}^{\dot{\gamma}k}
\epsilon_{\dot{\beta}\dot{\gamma}}\chi^{\alpha}_{ijk} +E^{c} \wedge  {E}{}^{\beta}_{ j} T_{\beta\;
c}^{j}{}\;^{\alpha}_{i} +E^{c} \wedge  \bar{E}{}^{\dot{\beta}j} T_{\dot{\beta}j\;
  c}{}\;^{\alpha}_{_{i}}
+\tfrac{1}{2} E^{c} \wedge E^{b}T_{bc}{}\; ^{\alpha}_{i}\; ,
\\
\label{Tfda=N8} T^{\dot{\alpha} i} & = & -\tfrac{1}{2} {E}{}^{{\beta}}_{j}\wedge {E}{}^{{\gamma}}_{k}
\epsilon_{{\beta}{\gamma}}\bar{\chi}{}^{\dot{\alpha}ijk} +E^{c} \wedge  {E}{}^{{\beta}}_{j} T^{j}_{{\beta}\;
c}{}^{\dot{\alpha} i} +E^{c} \wedge  \bar{E}{}^{\dot{\beta}j} T_{\dot{\beta}j\; c}{}^{\dot{\alpha}
  i}
+\tfrac{1}{2} E^{c} \wedge E^{b}T_{bc}{}^{\dot{\alpha} i}\; .\,\,\,\,\,\,\,\,\,\,\,\,
\end{eqnarray}
Here $\chi^{\alpha}_{ijk} =(\bar{\chi}{}^{\dot{\alpha}ijk})^{*}$ is the main fermionic superfield of
$\mathcal{N}=8,D=4$ supergravity and the dimension 1 fermionic torsion components have the expressions
\begin{equation}
  \begin{array}{rclrcl}
T_{\beta \, b}^{j}{}\; ^{\alpha}_{i} & = & \tfrac{1}{4}\chi_{ikl\beta}
(\bar{\chi}^{jkl}\tilde{\sigma}_{b})^{\alpha} \; , \hspace{2cm} & T_{\dot{\beta}j \, b}{}^{\dot{\alpha}i} & = &
\tfrac{1}{4}\bar{\chi}^{ikl}_{\dot{\beta}} (\tilde{\sigma}_{b}{\chi}_{jkl})^{\dot\alpha} \; ,
\\
T_{\dot{\beta}j \, b}{}\; ^{\alpha}_{i} & = & -\tfrac{i}{2} \sigma_{b\beta\dot{\beta}}  M^{\alpha\beta}_{ij} -
\tfrac{i}{2} \tilde{\sigma}_{b}^{\dot{\alpha}\alpha}\bar{N}_{\dot{\alpha}\dot{\beta}ij}\; , \hspace{.5cm}~ &
T_{{\beta} \, b}^{j}{}\; ^{\dot{\alpha} i} & = & -\tfrac{i}{2} \sigma_{b\beta\dot{\beta}}
\overline{M}^{\dot{\alpha}\dot{\beta}\, ij} - \tfrac{i}{2}
\tilde{\sigma}_{b}^{\dot{\alpha}\alpha}{N}_{{\alpha}{\beta}}^{i\; j} \; ,
\\
\end{array}
\end{equation}
in terms of the fermionic bilinears\footnote{$\varepsilon_{ij[3][3']}
\bar{\chi}_{\dot\alpha}^{[3]}\bar{\chi}_{\dot\beta}^{[3']}\equiv \varepsilon_{ijklmnpq}
\bar{\chi}_{\dot\alpha}^{klm}\bar{\chi}_{\dot\beta}^{npq}$}
\begin{equation}
\label{N=chch} {N}_{{\alpha}{\beta}}^{i\, j} = \frac{e^{-i\beta}}{6\cdot 4!}\varepsilon^{ij[3][3']}
\chi_{\alpha [3]}\chi_{\beta [3']}
 \; ,  \qquad
\bar{N}_{\dot{\alpha}\dot{\beta}ij} =
 -\frac{e^{i\beta}}{6\cdot 4!}\varepsilon_{ij[3][3']}
\bar{\chi}_{\dot\alpha}^{[3]}\bar{\chi}_{\dot\beta}^{[3']}
 \; ,
\end{equation}
and the bosonic superfields $M_{ij\alpha \beta} =M_{[ij](\alpha \beta )}=
(\bar{M}^{ij}_{\dot{\alpha}\dot{\beta}} )^{*}$. These appear as irreducible parts of the fermionic covariant
derivatives of the main fermionic superfield,
\begin{equation}
\label{Dch=M} D{}^{i}{}_{(\alpha} \chi_{\beta )jkl} = - 3\delta^{i}_{[j}M_{kl]\alpha \beta} \; , \quad
\bar{D}_{i(\dot{\alpha}} \bar{\chi}{}_{\dot{\beta})}{}^{jkl} = - 3
\delta_{i}^{[j}\bar{M}^{kl]}_{\dot{\alpha}\dot{\beta}} \; ,
\end{equation}
The other irreducible components of these covariant derivatives of the main superfield are expressed through
their bilinears:
\begin{equation}
\label{Dch=chch} D^{\alpha i} \chi_{\alpha jkl} = -\frac{e^{i\beta}}{12}\varepsilon_{jkl[2][3]}
\bar{\chi}_{\dot{\alpha}}^{i[2]}\bar{\chi}^{\dot{\alpha}[3]}\; , \qquad \bar{D}{}_{i}^{\dot{\alpha}}
\bar{\chi}{}_{\dot{\alpha}}^{jkl} = -\frac{e^{-i\beta}}{12}
\varepsilon^{jkl[2][3]}\chi^{\alpha}_{i[2]}\chi_{\alpha [3]} \; ,
\end{equation}

\subsection{$E_{7(+7)}/SU(8)$ Cartan forms in $\mathcal{N}=8$ supergravity
  superspace }

The covariantly constant Cartan 1-forms  obey the constraints
\begin{eqnarray}
\label{P=Echi+} \mathbb{P}_{ijkl} & =  & 2E^{\alpha }_{[i} \chi_{jkl]\alpha} -2
\frac{e^{i\beta}}{4!}\bar{E}^{\dot{\alpha}p} \varepsilon_{ijklp[3]}\bar{\chi}_{\dot{\alpha}}^{[3]}
+E^{a}\mathbb{P}_{a\; ijkl} \; , \\ \label{bP=bEchi+} \bar{\mathbb{P}}^{ijkl} & = &
2\frac{e^{-i\beta}}{4!}E^{\alpha }_{p} \varepsilon^{ijklp[3]}{\chi}_{{\alpha}[3]} -2
\bar{E}^{\dot{\alpha}[i}\bar{\chi}_{\dot{\alpha}}^{jkl]} +E^{a}\bar{\mathbb{P}}^{ijkl}_{a} \; .
\end{eqnarray}
These coincide with those in Refs.~\cite{Brink:1979nt,Howe:1981gz} up to the constant phase parameter $\beta$.
With the constraints (\ref{P=Echi+}), (\ref{bP=bEchi+}) the ``Bianchi identities'' (\ref{DP=0}) imply
 \begin{equation}
\label{bDchi=P} \bar{D}_{\dot{\alpha }i} \chi_{\alpha\, jkl} =
2i\sigma^{a}_{\alpha\dot{\alpha}}\mathbb{P}_{aijkl}\; , \hspace{1cm} D_{\alpha}^{i}
\bar{\chi}_{\dot{\alpha}}^{jkl} = - 2i\sigma^{a}_{\alpha\dot{\alpha}} \bar{\mathbb{P}}^{ijkl}_{a}\; ,
\end{equation}
The results of Eq.~(\ref{DP=0}) are also of help to find the expression Eq.~(\ref{Dch=chch}) for $D^{\alpha
i}\chi_{\alpha jkl}$, and the duality relation between the vector $\mathbb{P}_{aijkl}$ and its conjugate
$\bar{\mathbb{P}}_{a}^{ijkl}$
\begin{equation}
\label{P=*bP} \mathbb{P}_{aijkl} = \frac{e^{i\beta}}{4!} \varepsilon_{ijklpqrs}\bar{\mathbb{P}}_{a}^{pqrs}\; .
\end{equation}
Just after this stage the superspace 1-forms in Eq.~(\ref{P=Echi+}) and (\ref{bP=bEchi+}) become related by
Eq.~(\ref{bP=*P=eP}).

\section{1-form gauge potentials in $\mathcal{N}=8$ supergravity superspace}

Although the supervielbein forms restricted by the torsion constraints already contain all the fields of
supergravity multiplets, including the vector fields and their field strength, it is possible and also
convenient to introduce the corresponding 1-form gauge potentials in superspace. As it was found already in
Ref.~\cite{Brink:1979nt}, to preserve manifest $SU(8)$ R-symmetry, one should introduce the super-1-forms
corresponding to both the ``electric'' gauge fields of the supergravity multiplet and to their magnetic duals,
packed in the complex 1-form $A_{ij}= A_{[ij]}=dZ^{M} A_{M\, ij}(Z)$ in the {\bf 28} representation of $SU(8)$,
and its complex conjugate $\bar{A}^{ij}= \bar{A}^{[ij]}=dZ^{M} \bar{A}_{M}^{ ij}(Z)=(A_{ij})^{*} $ in its
$\overline{\bf 28}$ representation.

Their 2-form field strengths, which obey the generalized Bianchi identities (gBIs)
\begin{equation}
\label{DF=PbF} DF_{ij} = \mathbb{P}_{ijkl}\wedge \bar{F}^{kl}\; , \qquad D\bar{F}^{ij} =
\bar{\mathbb{P}}{}^{ijkl}\wedge {F}_{kl}\; ,
\end{equation}
are restricted by  the constraints
\begin{eqnarray}
\label{F=N8E7} F_{ij} & = & -i E^{\alpha}_{i} \wedge E^{\beta}_{j}\epsilon_{\alpha\beta}
-\tfrac{1}{2}E^{a}\wedge \bar{E}^{\dot{\gamma}k} \sigma_{a{\gamma}\dot{\gamma}}\chi^{\gamma}_{ijk}
+\tfrac{1}{2} E^{c}\wedge E^{b} F_{bc\; ij}\; ,
\\
\label{bF=N8E7}
 \bar{F}^{ij}
& = & -i E^{\dot{\alpha}i} \wedge E^{\dot{\beta}j}\epsilon_{\dot{\alpha}\dot{\beta}} +\tfrac{1}{2}E^{a}\wedge
{E}^{\gamma}_{k} \sigma_{a{\gamma}\dot{\gamma}}\bar{\chi}{}^{\dot{\gamma}ijk} +\tfrac{1}{2} E^{c}\wedge E^{b}
\bar{F}_{bc}^{i\,j}\; ,
\end{eqnarray}
The antisymmetric tensor superfield can be decomposed in the two irreducible parts\footnote{Notice that
$F_{\dot{\alpha}\dot{\beta}\; ij}= + \tfrac{1}{4}
  F_{ab\; ij}\tilde{\sigma}^{ab}_{\dot{\alpha}\dot{\beta}}= -
  (\bar{F}_{\alpha\beta}{}^{ij})^{*}$.}
\begin{equation}
\label{Fab:=} \sigma^{a}_{\alpha\dot\alpha} {\sigma}^{b}_{\beta \dot{\beta}}F_{ab\; ij} = 2
\epsilon_{\alpha\beta} F_{\dot{\alpha}\dot{\beta}\; ij} -2 \epsilon_{\dot{\alpha}\dot{\beta}}
 F_{\alpha\beta\; ij}\; .
\end{equation}
The Bianchi identities, including (\ref{DF=PbF}), imply, in particular,
\begin{equation}
\label{Fff=} F_{\alpha\beta\; ij} = \tfrac{i}{2} M_{\alpha\beta\; ij}\; , \qquad F_{\dot{\alpha}\dot{\beta}\;
ij} = \tfrac{i}{2} \bar{N}_{\dot{\alpha}\dot{\beta}\; ij} = -i {e^{i\beta}\over 12\cdot
4!}\varepsilon_{ij[3][3']} \bar{\chi}_{\dot{\alpha}}^{[3]}\bar{\chi}_{\dot{\beta}}^{[3']}\; .
\end{equation}


\section{2-form gauge potentials in  $\mathcal{N}=8$ supergravity superspace }

Now we are ready to turn to the main subject of this paper: 2-form gauge potentials $B_{2}^{\tilde{\Sigma}}$
(``notophs'') in the complete supersymmetric description of $\mathcal{N}=8, D=4$ supergravity.

As we have discussed in the introduction, although the appearance of seven 2-form potentials after dimensional
reduction from $D=11$ down to $D=4$ is manifest and was noticed already in \cite{Cremmer:1979up}, these were
immediately dualized to scalars. Only then the global $E_{7(+7)}$ duality becomes manifest.  The inverse
transformations relating all the scalars of $\mathcal{N}=8$ supergravity, parametrizing $E_{7(+7)}/SU(8)$, to
2-form potential have not been studied, at least in a complete form and especially in superspace, and this is
our goal here. As we are going to see, in addition to the 2-forms associated to the coset generators,
$B_{2}^{G/H}$, which were expected as dual to the physical scalars parametrizing $G/H=E_{7(+7)}/SU(8)$
(basically because there are 70 of them), it is necessary to introduce $\mathfrak{su}(8)$ valued 2-form
$B_{2}^{H}$. These are auxiliary and do not correspond to any dynamical degrees of freedom of
$\mathcal{N}=8,D=4$ supergravity. The general situation will be discussed in the companion paper
\cite{IB+TO=Next}. Here we adopt a more technical superspace-based approach to establishing the content and the
structure of the tensorial hierarchy of $\mathcal{N}=8, D=4$ supergravity.

\subsection{Strategy}

Our strategy to search for higher form potentials in maximal supergravity is essentially superspace based: we
begin by searching for an {\it ansatz} for possible superspace constraints for 3-form field strengths
${H}_{3}^{\tilde{\Sigma}}=dB_{2}^{\tilde{\Sigma}}+\ldots $ suggested by the indices carried by the potentials.
Checking their consistency, we can find whether more forms have to be introduced and what kind of ''free
differential algebra'' (FDA) they have to generate. This is described by a set of generalized Bianchi
identities (gBIs) $DF_{4}^{\mathcal{A}}=\ldots$ The further study of the gBIs (FDA relations) for the
constrained field strength should result (provided the constraints are consistent and the potentials are
dynamical fields) in equations of motion which, in the case of the 2-form potentials, should have the form of
duality of their field strength to the covariant derivatives of the scalar fields.  Since, in our case, these
are (the bosonic leading components of) the covariant $E_{7(7)}/SU(8)$ Cartan forms, i.e.~the complex self-dual
1-forms $P_{ijkl}= \tfrac{1}{4!} \varepsilon_{ijkli^{\prime}j^{\prime}k^{\prime}l^{\prime}}
\bar{P}{}^{i^{\prime}j^{\prime}k^{\prime}l^{\prime}}$ in the {\bf 70} of $SU(8)$ ($e^{-i\beta/2}\mathbb{P}_{a\,
ijkl}= \tfrac{1}{4!} \varepsilon_{ijklpqrs}e^{i\beta/2}\bar{\mathbb{P}}_{a}^{pqrs}$ in terms of bosonic
component of superforms), the ``physical'' 2-form potentials are expected to be $B_{2\, ijkl}$ and its complex
conjugate and dual $\bar{B}_{2}^{ijkl}$.

However, as discussed in the introduction, experience suggests that, when the scalars parametrize a coset space
$G/H$, it is not sufficient to introduce only the dual $(D-2)$-form potentials with indices of the generators
of the coset: the $(D-2)$-forms associated to the generators of the subgroup $H$ must be included as well (see
\cite{IB+TO=Next} for a general discussion).  In our case, these correspond to the hermitian traceless matrix
of 2-forms $B_{2
  i}{}^{j}= (B_{2 i}{}^{j})^{*}$ with the generalized field strength $H_{3\,
  i}{}^{j}=dB_{2\, i}{}^{j}+\ldots $.

\subsection{Constraints and generalized Bianchi identities for 3-form field
  strengths}

The natural candidate for the superspace constraints are
\begin{equation}
\label{H3ijkl=}
\begin{array}{rcl}
H_{3\; ijkl} & = & E^{\; \alpha}_{[i}\wedge \sigma^{(2)}{}_\alpha{}^{\beta}\chi_{ jkl] \beta} -
{\displaystyle\frac{e^{i\beta}}{4!}} \varepsilon_{ijkli^{\prime}j^{\prime}k^{\prime}l^{\prime}}
\bar{E}{}^{\dot{\alpha} i^{\prime}}\wedge \tilde{\sigma}{}^{(2)\; \dot{\beta}}{}_{\dot\alpha} \,
\bar{\chi}{}_{\dot{\beta}}{}^{ j^{\prime}k^{\prime}l^{\prime}} +
\\
& & \\ & &  + \tfrac{1}{3!} E^{c}\wedge E^{b}\wedge E^{a} H_{abc\; ijkl}\; ,
\end{array}
\end{equation}
where $\sigma^{(2)}{}_\alpha{}^{\beta}= {1\over 2}E^b\wedge E^a \sigma_{ab}{}_\alpha{}^{\beta} = -(\tilde{\sigma}{}^{(2)\; \dot{\beta}}{}_{\dot\alpha})^*$,
and
\begin{equation}
\label{H3ij=} H_{3\, i}{}^{j} = iE^{a} \wedge E^{\alpha}_{i} \wedge E^{\dot{\alpha}j}\sigma_{a
\alpha\dot{\alpha}} -\tfrac{i}{8} \delta_{i}{}^{j} \, E^{a} \wedge E^{\alpha}_{k} \wedge
E^{\dot{\alpha}k}\sigma_{a \alpha\dot{\alpha}} +\tfrac{1}{3!} E^{c} \wedge E^{b} \wedge E^{a} H_{abc\,
i}{}^{j}\; .
\end{equation}

Clearly, the leading term in the expression for $H_{3\; ijkl}$ should be $dB_{2\; ijkl}$. But the question to
be answered is whether other terms are also present, and the answer is affirmative. Indeed, if we assume
$H_{3\;
  ijkl}=dB_{2\; ijkl}$ (or, keeping the $SU(8)$ invariance, $H_{3\;
  ijkl}=DB_{2\; ijkl}$), the generalized field strength should obey the
simplest Bianchi identities $dH_{3\; ijkl}=0$ (or $DH_{3\; ijkl}=4R_{[i|}{}^p \wedge H_{3\; |jkl]p}$), and the
constraints (\ref{H3ijkl=}) are not consistent if consistency is expressed by such a simple Bianchi identity.

Similarly one can check that no consistent FDA can be formulated without introducing also the
$\mathfrak{su}(8)$ valued field strength ${H}_{3i}{}^{j}$.  It might also look tempting to omit the
tracelessness condition $H_{3\, i}{}^{i}=0$ and thus to consider the $\mathfrak{u}(8)$ rather than
$\mathfrak{su}(8)$ valued 3-form field strength obeying simpler constraints given by (\ref{H3ij=}) without the
second term in the r.h.s. However, as we have checked, this is also inconsistent with the superspace
constraints of $\mathcal{N}=8$ supergravity. Thus the structure of the tensor hierarchy of $\mathcal{N}=8$
supergravity is quite rigid.

To make a long story short, we have found that the constraints (\ref{H3ijkl=}) and (\ref{H3ij=}) are consistent with the FDA relations (generalized Bianchi identities)
\begin{equation}
\label{I4ij=DHij} I_{4\, i}{}^{j} := DH_{3\, i}{}^{j} +2 F_{ik}\wedge \bar{F}^{jk}
-\tfrac{1}{4}\delta_{i}{}^{j} F_{kl}\wedge \bar{F}^{kl} +\tfrac{1}{3} H_{3ikpq} \wedge
\bar{\mathbb{P}}{}^{jkpq} +\tfrac{1}{3} \bar{H}_{3}^{jkpq} \wedge \mathbb{P}_{ikpq}=0\; , \qquad
\end{equation}
and
\begin{eqnarray}
\label{I43=DH3} I_{4\; ijkl} := DH_{3\; ijkl} -4 {H}_{3\, [i}{}^{j^{\prime}} \wedge \mathbb{P}_{jkl]j^{\prime}}
-3 F_{[ij}\wedge F_{kl]} +\frac{3e^{i\beta}}{4!}\varepsilon_{ijkli^{\prime}j^{\prime}k^{\prime}l^{\prime}}
\bar{F}{}^{i^{\prime}j^{\prime}}\wedge \bar{F}^{k^{\prime}l^{\prime}} =0\; . \qquad
\end{eqnarray}

Let us stress that
\begin{enumerate}
\item As long as $\tilde{H}_{3 \, p}{}^{[i }\wedge
  \bar{\mathbb{P}}{}^{jkl]p}=- {e^{-i\beta}\over
    4!}\varepsilon^{ijkli^{\prime}j^{\prime}k^{\prime}l^{\prime}} \tilde{H}_{3 \,
    i^{\prime}}{}^{p }\wedge \mathbb{P}_{j^{\prime}k^{\prime}l^{\prime}p}$,
  the identity (\ref{I43=DH3}) and the complex conjugate identity for
  $\bar{H}_{3}{}^{ijkl}=(H_{3\; ijkl})^{*}$ are consistent with the duality
  relation ({\it cf.} with (\ref{bP=*P=eP}); notice the sign)
\begin{equation}
\label{DtH3ij=FbF+} \bar{H}_{3}^{ijkl} = - \frac{e^{-i\beta}}{4!}
\varepsilon^{ijkli^{\prime}j^{\prime}k^{\prime}l^{\prime}} H_{3\; i^{\prime}j^{\prime}k^{\prime}l^{\prime}}
\; ,  \qquad
\end{equation}
\item When this property is taken into account, the traces of last two terms
  in the {\it r.h.s.} of (\ref{I4ij=DHij}) cancel one another.
\item The terms quadratic in 2-form field strengths are those that occur in
  the $E_{7(+7)}$ Noether-Gaillard-Zumino current \cite{Gaillard:1981rj}. This
  current, whose components are all conserved, even for the $E_{7(+7)}$
  transformations which are not symmetries of the action, may play an
  important role in the UV finiteness of the theory \cite{Brodel:2009hu}.
\end{enumerate}

To check the consistency of our ansatz for the generalized Bianchi identities (gBIs) (\ref{I4ij=DHij}) and
(\ref{I43=DH3}) one has to study the ''identities for identities'' $I_{5}^{G}=DI_{4}^{G}=0$:
\begin{eqnarray}
\label{idFORid=0} I_{5\, i}{}^{j} := DI_{4\, i}{}^{j}= 0 \; , \qquad I_{5 ijkl} := DI_{4 ijkl}=0 \; ,
\end{eqnarray}
taking into account the Ricci identities. In application to our 3-form the latter read
\begin{equation}
\label{DDH=RH} DD{H}_{3\, i}{}^{j} = R_{i}{}^p\wedge  {H}_{3\, p}{}^{j} -{H}_{3\, i}{}^p\wedge  R_{p}{}^{j} \,
, \qquad  DDH_{3 ijkl}=4R_{[i|}{}^p\wedge H_{3 p|jkl]}\; ,
\end{equation}
and can be further specified substituting the explicit expression (\ref{Rij=PP}) for the curvature of induced
$SU(8)$ connection. In such a way, after some algebra, one can prove that the proposed gBIs (\ref{I4ij=DHij})
and (\ref{I43=DH3}) are consistent provided the following identity holds
\begin{equation}
\label{PbPH=id} \mathbb{P}_{[3][i|}\wedge \bar{\mathbb{P}}{}^{[3]q}\wedge H_{3\, |jkl]q} -
\mathbb{P}_{p[ijk|}\wedge \bar{\mathbb{P}}{}^{p[3]}\wedge H_{3\, |l][3]} - \mathbb{P}_{p[ijk}\wedge
\mathbb{P}_{l][3]}\wedge \bar{H}_{3}{}^{p[3]} =0 \; .
\end{equation}
This equation is proven in Appendix~\ref{sec-onE77} using only the complex self-duality and anti-self-duality
of $\mathbb{P}_{ijkl}$ and $H_{3\, ijkl}$, respectively.

\subsection{Superfield duality equations}
\label{dualitySec} 

Substituting Eqs.~(\ref{H3ij=}) and (\ref{H3ijkl=}) and using the superspace supergravity constraints, we have
checked that the dim 2 and 5/2 components of the gBIs (\ref{I43=DH3}) and (\ref{I4ij=DHij}) are satisfied. As
far as dim 3 components are concerned, the $\propto E^{b}\wedge E^{a} \wedge E^{\alpha}_{p}\wedge
E^{\beta}_{q}$ component of Eq.~(\ref{I43=DH3}) is satisfied identically (due to the basic constraints and
properties of main superfields, like (\ref{Dch=M}) with (\ref{Fff=})), while its $\propto E^{b}\wedge E^{a}
\wedge E^{\alpha}_{p}\wedge \bar{E}^{\dot{\beta}q}$ component shows that $H_{abc\, ijkl}$ is dual to the
generalized Cartan form $\mathbb{P}^{d}_{ijkl}$,
\begin{equation}
\label{Habc4=eP4} H_{abc\; ijkl} = \tfrac{i}{2} \epsilon_{abcd} \mathbb{P}^{d}_{ijkl} \; .
\end{equation}
The $\propto E^{b}\wedge E^{a} \wedge E^{\alpha}_{p}\wedge \bar{E}^{\dot{\beta}q}$ component of (\ref{I43=DH3})
shows that $H_{abc}{}_{i}{}^{j}$ is dual to a bilinear of fermionic superfields,
\begin{eqnarray}
\label{Habc2=ff} H_{abc}{}_{i}{}^{j} \propto \epsilon_{abcd} \left( \chi_{i[2]} \sigma^{d} \bar{\chi}^{j[2]}
-\tfrac{1}{8}\delta_{i}^{j} \chi_{[3]} \sigma^{d} \bar{\chi}^{[3]}\right)\; .
\end{eqnarray}
This reflects the auxiliary character of the $\mathfrak{su}(8)$ ``(pseudo-)notophs''.

\subsection{Identities  for identities and the  proof of the consistency of the constraints}
\label{idForBID} 

Instead of studying the higher-dimensional components of the gBIs, we simplify our study by proving that they
are dependent and cannot produce independent consequences; this implies that our constraints are consistent and
all the dynamical equations are contained as higher components in the superfield duality equations
(\ref{Habc4=eP4}) and (\ref{Habc2=ff}).

To this end we solve the identities for identities (\ref{idFORid=0}), $0= I_{5}^{G}=(I_{5\, ijkl}, I_{5\,
i}{}^{j})=DI_{4}^{G}$, with respect to the (l.h.s.~of the) gBIs, $I_{ABCD}^{G}$, in the same manner as we solve
Bianchi identities for the torsion and curvature tensors (and also gBIs for the 3-forms above) expressing them
in terms of the main superfields (see Ref.~\cite{Sohnius:1980iw}).

As we have already said, the lower dimensional, dim 2 and 5/, components of the 4-form gBIs are satisfied
algebraically, without any involvement of superfields. Setting these to zero,
$I^{G}_{\underline{\alpha}\underline{\beta}\underline{\gamma} A}=0$, we obtain a counterpart of the torsion
constraints of supergravity. Substituting
\begin{equation}
  \begin{array}{rcl}
\label{IG=dim3+} I_{4}^{G} & = & \tfrac{1}{4}E^{b} \wedge E^{a} \wedge E^{\underline{\alpha}} \wedge
E^{\underline{\beta}}\, I_{  \underline{\beta}\underline{\alpha}ab}^{G} +\tfrac{1}{3!}E^{c} \wedge E^{b} \wedge
E^{a} \wedge E^{\underline{\alpha}}  \, I_{ \underline{\alpha}abc}^{G}
 \\
& & \\ & & + \tfrac{1}{4!}E^{d} \wedge E^{c} \wedge E^{b} \wedge E^{a} I_{ abcd}^{G}\; ,
\end{array}
\end{equation}
into Eq.~(\ref{idFORid=0}) and using the torsion constraints of $\mathcal{N}=8,D=4$ supergravity,
Eqs.~(\ref{Ta=N8}), (\ref{Tfa=N8}) and (\ref{Tfda=N8}), we find
\begin{equation}
\label{IG5=bffff+} 0 = I_{5}^{G} = -\tfrac{i}{2}E^{b} \wedge E^{\alpha}_{p} \wedge E^{\dot{\alpha}q} \wedge
E^{\underline{\beta}} \wedge E^{\underline{\gamma}}\, \delta_{q}^{p}\, \sigma^{a}_{\alpha\dot{\alpha}}\,
I^{G}_{\underline{\beta}\, \underline{\gamma}\, ab} + \propto  E^{b} \wedge E^{a}
 \qquad
\end{equation}
Thus, the lowest dimensional (dim 3) nontrivial components of the identities for identities imply the following
algebraic equations for the l.h.s.~of the dim 3 gBIs:
\begin{eqnarray}
\label{IG5=d3-1} 0 & = & \delta_{q}^{p}\sigma^{a}_{ \alpha\dot{\alpha}} I^{G}{}^{k}_{ {\beta}\, \dot{\gamma}
l\, ab} + \delta_{q}^{k}\, \sigma^{a}_{{\beta}\dot{\alpha}} I^{G}{}^{p}_{ \alpha\, \dot{\gamma} l\, ab} +\left(
{\dot{\alpha}q} \; \mapsto\; {\dot{\gamma}l} \right)\; ,
\\
& & \nonumber \\ \label{IG5=d3-2} 0 & = & \delta_{q}^{p}\, \sigma^{a}_{ \alpha\dot{\alpha}} I^{G}{}^{k\, l}_{
{\beta}\, {\gamma} \, ab} +\delta_{q}^{k}\, \sigma^{a}_{ \beta\dot{\alpha}} I^{G}{}^{l\, p}_{{\gamma}\,
{\alpha} \, ab} +\delta_{q}^{l}\, \sigma^{a}_{ \gamma\dot{\alpha}} I^{G}{}^{p\, k}_{\alpha \, {\beta} \, ab}\;
,
\end{eqnarray}
plus the complex conjugate of Eq.~(\ref{IG5=d3-2}). It is not difficult to find that the latter as well as
Eq.~(\ref{IG5=d3-2}) have only trivial solutions $I{}^{G}{}^{k\, l}_{ {\beta}\, {\gamma} \, ab} =0$. In
contrast, the general solution of Eq. (\ref{IG5=d3-1}) reads $I^{G}{}_{\alpha\dot{\alpha} j
  bc}^{i}{}= \delta_{j}^{i}\,
\sigma^{a}_{\alpha\dot{\alpha}}\tilde{I}_{abc}^{G}$ with an arbitrary antisymmetric
$\tilde{I}_{abc}^{G}=\tilde{I}_{[abc]}^{G}$. This implies that the only independent consequences for the
superfields can be obtained from $I^{G}{}_{\alpha\dot{\alpha} j
  [bc}^{j}\tilde{\sigma}_{a]}^{\alpha\dot{\alpha}}=0$.

This is exactly what we have observed in the explicit calculations of the dimension 3 Bianchi identities for
$H_{3\, ijkl}$ (see Sec.~\ref{dualitySec}). Namely, we have found that
\begin{eqnarray}
\label{I4dim3=} 0 = (I_{4\, ijkl}){}_{\alpha\dot{\alpha} q\, ab}^{p}\, \equiv\, -i \delta_{q}^{p}\,
\sigma^{c}_{\alpha\dot{\alpha}}  \left(H_{abc\; ijkl}-  \tfrac{i}{2} \epsilon_{abcd}
\mathbb{P}^{d}_{ijkl} \right)\; ,
\end{eqnarray}
which implies the superfield duality equation (\ref{Habc4=eP4}).

The above general statement allows one to escape the exhausting algebraic calculations necessary to check
explicitly the cancellation of different terms in the equation $I^{G}{}_{\underline{\alpha}\underline{\beta} ab}=0$.

Furthermore, the higher-dimensional components of identities for identities Eq.~(\ref{IG5=bffff+}) show the
dependence of higher-dimensional Bianchi identities $I_{\underline{\alpha}abc}^{G} =0$ and $I_{ abcd}^{G}=0$.
This implies that their results can be obtained by applying covariant derivatives to the results of the
dimension $3$ gBIs, this is to say to the superfield duality equations (\ref{Habc4=eP4}) and (\ref{Habc2=ff}),
with the use of the superspace constraints for torsion, Cartan forms and 2-form field strength of the 1-form
gauge fields and of their consequences. The latter include the equations of motion of $\mathcal{N}=8,D=4$ CJ
supergravity.

\subsection{Scalar (super)field equation of motion and duality equation}

To illustrate this statement let us consider the dimension 4 Bianchi identity corresponding to
$E_{7(+7)}/SU(8)$ generators:
\begin{eqnarray}
\label{I4dim4=} 0 = I_{ijkl\, abcd} & = & 4D_{[a} H_{bcd] \, ijkl} +16 H_{[abc|[i}{}^{p}\mathbb{P}_{jkl]p\,
|d]} -18 F_{[ij|[ab} F_{cd]|kl]} \nonumber \\ & & +\frac{3e^{i\beta}}{4}
\varepsilon_{ijkli^{\prime}j^{\prime}k^{\prime}l^{\prime}} \bar{F}_{[ab}^{i^{\prime}j^{\prime}}
\bar{F}_{cd]}^{\prime k^{\prime}l^{\prime}} +6 T_{[ab|}{}^{\alpha}_{[i|} (\sigma_{|cd]}\chi_{|jkl]})_{\alpha}
\nonumber \\ & & -\frac{e^{i\beta}}{4} \varepsilon_{ijkli^{\prime}j^{\prime}k^{\prime}l^{\prime}}
T_{[ab|}{}^{\dot{\alpha}i^{\prime}} (\bar{\chi}{}^{j^{\prime}k^{\prime}l^{\prime}}
\tilde{\sigma}_{|cd]})_{\dot\alpha}\; .
\end{eqnarray}
Using (\ref{Habc4=eP4}) we can equivalently write this as
\begin{eqnarray}
\label{DaPa4=H+} D^{a} \mathbb{P}_{a\, ijkl} & = & -\tfrac{4i}{3} \varepsilon^{abcd}
H_{abc[i}{}^{p}\mathbb{P}_{jkl]p\, d} -\tfrac{3i}{2} \varepsilon_{abcd} F_{[ij}^{ab} F_{kl]}^{cd}
+\frac{ie^{i\beta}}{16} \varepsilon_{ijkli^{\prime}j^{\prime}k^{\prime}l^{\prime}} \varepsilon^{abcd}
\bar{F}_{ab}^{i^{\prime}j} \bar{F}_{cd}^{'k^{\prime}l^{\prime}} \nonumber \\ & & +T_{ab}{}^{\alpha}_{[i}
(\sigma^{ab}\chi_{jkl]})_{\alpha} +\frac{e^{i\beta}}{4!}
\varepsilon_{ijkli^{\prime}j^{\prime}k^{\prime}l^{\prime}} T_{ab}{}^{\dot{\alpha}i^{\prime}}
(\bar{\chi}^{j^{\prime}k^{\prime}l^{\prime}}\tilde{\sigma}^{ab})_{\dot\alpha} \; .
\end{eqnarray}
After using Eq.~(\ref{Habc2=ff}), this expression acquires the usual form of the scalar (super)field equation of
$\mathcal{N}=8,D=4$ supergravity,
\begin{equation}
\label{DaPa4=Eq} D^{a} \mathbb{P}_{a\, ijkl} = -\tfrac{3i}{2} \varepsilon_{abcd} F_{[ij}^{ab} F_{kl]}^{cd}
+\frac{ie^{i\beta}}{16} \varepsilon_{ijkli^{\prime}j^{\prime}k^{\prime}l^{\prime}} \varepsilon^{abcd}
\bar{F}_{ab}^{i^{\prime}j^{\prime}} \bar{F}_{cd}^{\prime  k^{\prime}l^{\prime}} +\ldots \; , \qquad
\end{equation}
where the dots stand for the terms bilinear in fermions.

To reflect the dependence of the higher-dimensional Bianchi identities proved in the previous
Sec.~\ref{idForBID}, the above line should be read in the opposite direction: the results of the dimension 4
Bianchi identity Eq.~(\ref{I4dim4=}) can be obtained by taking the bosonic covariant derivative of the duality
equation (\ref{Habc4=eP4}) and using the scalar (super)field equation (as obtained from the torsion constraints
of \cite{Brink:1979nt,Howe:1981gz}) and Eq.~(\ref{Habc2=ff}).

Thus, the results of Sec.~\ref{dualitySec} and the arguments of Sec.~\ref{idForBID} allow us to conclude that
our constraints for the 3-form field strength are consistent and describe a set of ``notophs'' dual to the
scalar fields of $\mathcal{N}=8,D=4$ supergravity.

\section{Conclusion and outlook}

In this paper we have provided the complete supersymmetric description of the ``notophs'' (2-form gauge
potentials) of the Cremmer-Julia $\mathcal{N}=8,D=4$ supergravity \cite{Cremmer:1979up}. More specifically, we
have presented the set of superspace constraints for the 3-form field strengths of the 2-form gauge potentials
defined on $\mathcal{N}=8,D=4$ supergravity superspace \cite{Brink:1979nt} and we have shown that these are
consistent and produce the duality relation between the field strengths of the ``physical notophs'' and the
scalar fields of the $\mathcal{N}=8,D=4$ CJ supergravity parametrizing the $G/H=E_{(7(+7)}/SU(8)$ coset. We
have found that the consistency, expressed by the generalized Bianchi identities, requires to introduce also
the auxiliary 2-form potentials corresponding to the generators of the stability subgroup $H=SU(8)$ of the
coset. In the companion paper \cite{IB+TO=Next} we will discuss the reasons for this in detail. Here we have
adopted a purely superspace approach and arrived at this conclusion starting from the natural candidate for the
superspace constraints and searching for their consistency.  The generalized Bianchi identities for the 3-form
field strengths of the notophs, which define the tensorial hierarchy (or free differential algebra) of the
$\mathcal{N}=8,D=4$ CJ supergravity, have been also obtained in this manner.

The list of natural directions of development of our approach includes the studies of the superfield
description of the notophs of gauged $\mathcal{N}=8,D=4$ supergravity \cite{de Wit:1981eq,de
  Wit:1982ig,deWit:2008ta} using the torsion constraints of \cite{Howe:1981tp}
and of the supersymmetric aspects of the generalized ``notophs'' of the exceptional field theories
\cite{Hohm:2013pua,Hohm:2013uia,Godazgar:2014nqa} in $\mathcal{N}=8,D=4$ superspace enlarged by 56 bosonic
``central charge'' coordinates (see \cite{Howe:1980th}). Another obvious extension of this work is the search
for worldvolume actions of possible superstring models carrying the ``electric'' charges with respect to the
antisymmetric tensor gauge fields. Probably the correct posing of this problem may also require to work in
the Howe-Linmdst\"om enlarged $\mathcal{N}=8,D=4$ superspace.

\section*{Notice added}

After this paper appeared on the net, the superspace description of higher form gauge fields in D-dimensional maximal and half maximal supergravities have been discussed in \cite{Howe:2015hpa}, where the cases of D=11 and D=10 are elaborated explicitly. For $3\leq D<10$ cases the representations carried by higher forms in maximal and half maximal superspaces  and their generalized Bianchi identities have been tabulated in Appendix A of   \cite{Howe:2015hpa}.  Higher forms in maximal and half-maximal D=3 dimensional superspaces were studied in  \cite{Greitz:2011vh}.

\section*{Acknowledgments}

This work has been supported in part by the Spanish MINECO grants partially financed  with FEDER funds: No
FPA2012-35043-C02-01, the Centro de Excelencia Severo Ochoa Program grant SEV-2012-0249  and the Spanish
Consolider-Ingenio 2010 program CPAN CSD2007-00042, as well as  by the Basque Government research group grant ITT559-10 and the  Basque Country University program UFI 11/55. We are thankful to the Theoretical Department of CERN for hospitality and support of our visits in 2013, when the present project has been initiated. T.O. wishes to thank M.M.~Fern\'andez for her permanent
support.

\appendix
\section{4D Weyl spinors and sigma matrices}

We use the relativistic Pauli matrices $\sigma^{a}_{\beta\dot{\alpha}} = \epsilon_{\beta\alpha}
\epsilon_{\dot{\alpha}\dot{\beta}} \tilde{\sigma}{}^{a\dot{\beta}\alpha}$ which obey
 \begin{equation}
\label{sasb=}
 \sigma^{a}\tilde{\sigma}{}^{b}
= \eta^{ab} +\tfrac{i}{2}\epsilon^{abcd}\sigma_{c}\tilde{\sigma}_d\; , \qquad \tilde{\sigma}{}^{a}{\sigma}^{b}
= \eta^{ab} -\tfrac{i}{2}\epsilon^{abcd}\tilde{\sigma}_{c}{\sigma}_d\; ,
 \qquad
\end{equation}
where $\eta^{ab} = \mathrm{diag}(1,-1,-1,-1)$ is the Minkowski metric and $\epsilon^{abcd}=\epsilon^{[abcd]}$
is the antisymmetric tensor with $\epsilon^{0123}=1=-\epsilon_{0123}$.

The spinorial ($SL(2,\mathbb{C})$) indices are raised and lowered by $\epsilon^{\alpha\beta} =
-\epsilon^{\beta\alpha} = i\tau_{2} = \left(\begin{smallmatrix} 0 & 1 \\ -1 & 0 \end{smallmatrix}\right)=
-\epsilon_{\alpha\beta}$ obeying $\epsilon_{\alpha\beta}\epsilon^{\beta\gamma} = \delta_{\alpha}^{\gamma}$:
$\theta_{\alpha} =\epsilon_{\alpha\beta}\theta^{\beta}$ and $\theta^{\alpha}
=\epsilon^{\alpha\beta}\theta_\beta$.  The antisymmetrized products $\sigma^{ab}{}_{\beta} {}^{\alpha}
=\sigma^{[a}\tilde{\sigma}^{b]} := \tfrac{1}{2}(\sigma^{a}\tilde{\sigma}^{b}- \sigma^{b}\tilde{\sigma}^{a}) $
and $\tilde{\sigma}{}^{ab}{}^{\dot\alpha}{}_{\dot\beta}=\tilde{\sigma}^{[a}{\sigma}^{b]} $ are self-dual and
anti-self-dual, $\sigma^{ab}=\tfrac{i}{2}\epsilon^{abcd}\sigma_{cd}$, $\tilde{\sigma}{}^{ab}
=-\tfrac{i}{2}\epsilon^{abcd}\tilde{\sigma}_{cd}$.

\section{More on differential forms in curved $\mathcal{N}=8, D=4$ superspace}
\label{app-moreonforms}
\subsection*{Exterior derivative}

The exterior derivative $d$ acts on a q-form
\begin{displaymath}
\Omega_{q} = \tfrac{1}{q!}dZ^{M_{q}}\wedge \ldots \wedge dZ^{M_1} \Omega_{M_1 \ldots M_{q}}(Z)=
\tfrac{1}{q!}E^{A_{q}}\wedge \ldots \wedge E^{A_1} \Omega_{A_1 \ldots
  A_{q}}(Z)
\end{displaymath}
as
\begin{equation}
\begin{array}{rcl}
\label{dOm:=} d\Omega_{q} & = & \tfrac{1}{q!}dZ^{M_{q}}\wedge \ldots \wedge dZ^{M_{1}}
 \wedge  d\Omega_{M_{1} \ldots M_{q}}(Z)
\\
& & \\ & = & \tfrac{1}{(q+1)!}dZ^{M_{q+1}}\wedge \ldots \wedge dZ^{M_{2}} \wedge dZ^{M_{1}}\,
(q+1)\partial_{[M_{1}}\Omega_{M_{2} \ldots M_{q+1}\}}(Z) \; .
\end{array}
\end{equation}
In action on the product of differential forms, e.g.~the q-form $\Omega_{q}$ and the p-form $\Omega_{p}$, it
obeys the Leibnitz rule
\begin{equation}
\label{dOmOm=} d(\Omega_{q}\wedge \Omega_{p})
 =
\Omega_{q}\wedge d\Omega_{p} + (-)^{p} d\Omega_{q}\wedge \Omega_{p}\; .
\end{equation}
The mixed brackets $[\ldots \}$ denote the graded antisymmetrization of the enclosed indices with the weight
unity, so that $(q+1)! \partial_{[M_{1}}\Omega_{M_{2} \ldots M_{q+1}\}}(Z) = \partial_{M_{1}}\Omega_{M_{2} \ldots
M_{q+1}}(Z)-(-)^{\varepsilon (M_{1})
  \varepsilon (M_{2}) } \partial_{M_{2}}\Omega_{M_{1}M_{3} \ldots
  M_{q+1}}(Z)+\ldots $, where $\varepsilon (M):=\varepsilon (Z^{M})$ is the
Grassmann parity (fermionic number), $\varepsilon (\mu):=\varepsilon (x^{\mu}) =0$, $\varepsilon
(\underline{\alpha})=\varepsilon (\theta^{\underline{\alpha}})=1$.

\subsection*{On $E_{7(+7)}$ Cartan forms}
\label{sec-onE77}

Using the complex self-duality of $\mathbb{P}_{ijkl}$ Eq.~(\ref{bP=*P=eP}) and the antisymmetry of the exterior
product of $\mathbb{P}$ one finds
\begin{equation}
\label{P4bP4=0} \mathbb{P}_{[4]} \wedge \bar{\mathbb{P}}{}^{[4]} = 0\; .
\end{equation}
Then, using Eq.~(\ref{bP=*P=eP}) and this last property Eq.~(\ref{P4bP4=0}) in $\mathbb{P}_{ij[2]} \wedge
\bar{\mathbb{P}}{}^{kl[2]}$ one finds
\begin{equation}
\label{PijbPkl=} \mathbb{P}_{ijpq} \wedge \bar{\mathbb{P}}{}^{klpq} = \tfrac{2}{3}
\delta_{[i}{}^{[k}\mathbb{P}_{j][3]} \wedge \bar{\mathbb{P}}{}^{l][3]}\; .
\end{equation}
The Ricci identity
\begin{equation}
\label{PbPP=0} DD \mathbb{P}_{ijkl}=-4 R_{[i}{}^{p} \wedge \mathbb{P}_{jkl]p} = -\tfrac{4}{3} \mathbb{P}_{[3][i}\wedge
\bar{\mathbb{P}}{}^{[3]p} \wedge \mathbb{P}_{jkl]p} =0\;
\end{equation}
is satisfied because, by virtue of Eq.~ (\ref{PijbPkl=}), the r.h.s. is equivalent to
\begin{eqnarray}
\label{PbPP2=0} \mathbb{P}_{pq[ij}\wedge \bar{\mathbb{P}}{}^{pqrs} \wedge \mathbb{P}_{kl]rs} \; ,
\end{eqnarray}
which vanishes automatically on account of the antisymmetry of the wedge product and the symmetry under the
interchange of pairs of the $SU(8)$ indices.

From Eq.~(\ref{PijbPkl=}) it follows that the first term in Eq.~(\ref{PbPH=id}) can be reexpressed as
\begin{equation}
\label{eq:firstterm} \mathbb{P}_{[3][i|}\wedge \bar{\mathbb{P}}{}^{[3]q}\wedge H_{3\, |jkl]q} = -\tfrac{3}{2}
\mathbb{P}_{[2][ij|}\wedge \bar{\mathbb{P}}{}^{[2][2^{\prime}]}\wedge H_{3\,
  |kl][2^{\prime}]}\; .
\end{equation}
Using again the complex self-duality of $\mathbb{P}_{ijkl}$ and the complex anti-self-duality of $H_{3\,
ijkl}$, the third term in Eq.~(\ref{PbPH=id}) can be reexpressed as
\begin{equation}
\label{eq:thirdterm} \mathbb{P}_{p[ijk}\wedge \mathbb{P}_{l][3]}\wedge \bar{H}_{3}{}^{p[3]} = -\tfrac{1}{8}
\mathbb{P}_{ijkl}\wedge \mathbb{P}_{[4]}\wedge H_{3}{}^{[4]} -\tfrac{3}{4} \mathbb{P}_{[2][ij|}\wedge
\bar{\mathbb{P}}{}^{[2][2^{\prime}]}\wedge H_{3\,
  |kl][2^{\prime}]}
\; .
\end{equation}
The same properties and this last identity allow us to rewrite the second term in Eq.~(\ref{PbPH=id}) as
\begin{equation}
\label{eq:secondterm} \mathbb{P}_{p[ijk|}\wedge \bar{\mathbb{P}}{}^{p[3]}\wedge H_{3\, |l][3]} = \tfrac{1}{8}
\mathbb{P}_{ijkl}\wedge \mathbb{P}_{[4]}\wedge H_{3}{}^{[4]} -\tfrac{3}{4} \mathbb{P}_{[2][ij|}\wedge
\bar{\mathbb{P}}{}^{[2][2^{\prime}]}\wedge H_{3\,
  |kl][2^{\prime}]}
\; .
\end{equation}
After rewriting the three terms of Eq.~(\ref{PbPH=id}) using the above identities, we find that
Eq.~(\ref{PbPH=id}) is identically satisfied.

\subsection*{On $E_{7(+7)}$ Cartan forms in  $\mathcal{N}=8$ supergravity
  superspace}

Eqs.~(\ref{Dch=chch}) can be derived also from (\ref{DP=0}) with (\ref{P=Echi+}). To this end it is useful to
notice the trivial identity
\begin{equation}
\bar{\chi}^{\dot{\alpha} pq[i}\bar{\chi}_{\dot\alpha}^{jkl]} = \tfrac{5}{2}\bar{\chi}^{\dot{\alpha}
p[qi}\bar{\chi}_{\dot\alpha}^{jkl]} -\tfrac{3}{2}\bar{\chi}^{\dot{\alpha} p[ij}\bar{\chi}_{\dot\alpha}^{kl]q}\:
.
\end{equation}
Its l.h.s.~is antisymmetric, while the second term in its r.h.s~is symmetric. Hence
\begin{equation}
\bar{\chi}^{\dot{\alpha} p[ij}\bar{\chi}_{\dot\alpha}^{kl]q} = \tfrac{5}{6} (\bar{\chi}^{\dot{\alpha}
p[qi}\bar{\chi}_{\dot\alpha}^{jkl]} +\bar{\chi}^{\dot{\alpha} q[pi}\bar{\chi}_{\dot\alpha}^{jkl]} )\; ,
\end{equation}
and
\begin{equation}
\bar{\chi}^{\dot{\alpha} pq[i}\bar{\chi}_{\dot\alpha}^{jkl]} = \tfrac{5}{4}(\bar{\chi}^{\dot{\alpha}
p[qi}\bar{\chi}_{\dot\alpha}^{jkl]} -\bar{\chi}^{\dot{\alpha} q[pi}\bar{\chi}_{\dot\alpha}^{jkl]} )\; .
\end{equation}
As a consequence
\begin{equation}
\varepsilon^{ijkli^{\prime}j^{\prime}k^{\prime}l^{\prime}}
\bar{\chi}{}^{\dot\alpha}_{pq[i^{\prime}}\chi_{\dot{\alpha}[j^{\prime}k^{\prime}l^{\prime}]} =
-2\delta_{[i}^{[p} \varepsilon^{jkl[2][3]} \bar{\chi}{}^{\dot\alpha}_{q][2]} \chi_{\dot{\alpha}[3]}\; .
\end{equation}
\subsection*{Curvature 2 forms of $\mathcal{N}=8$ superspace}

The study of the Bianchi identities results in the following expressions for the curvature of the spin
connection (the ``Riemann'' curvature 2-form) (see \cite{Brink:1979nt}, \cite{Howe:1981gz})
\begin{eqnarray}
\label{stsR=} \sigma^{a}_{\alpha\dot\alpha} \tilde{\sigma}_{b}^{\dot{\beta}\beta}R_{a}{}^{b} & = &
2\delta_{\alpha}{}^{\beta} R_{\dot\alpha}{}^{\dot{\beta}} +2 \delta_{\dot\alpha}{}^{\dot{\beta}}
R_{\alpha}{}^{\beta} \; , \\ & & \nonumber \\ \label{Rff=N8} R^{\alpha\beta} =
\tfrac{1}{4}R^{ab}\sigma_{ab}^{\alpha\beta} & = & \tfrac{1}{2} {E}{}^{{\gamma}}_{i}\wedge
{E}{}^{{\delta}}_{j}\left(\epsilon_{\gamma\delta} N^{\alpha\beta ij}+
2\delta_{(\gamma}{}^{\alpha}\delta_{\delta )}{}^{\beta} S^{ij}\right) +\tfrac{1}{2} \bar{E}{}^{\dot{\beta}i}
\wedge  \bar{E}{}^{\dot{\gamma}j} \epsilon_{\dot{\gamma} \dot{\delta}} M^{\alpha\beta}_{ij} \nonumber \\ & &
\nonumber \\ & & +{E}{}^{{\gamma}}_{i}\wedge \bar{E}{}^{\dot{\gamma}j} R^{i}_{{\gamma}\;\dot{\gamma}j
}{}^{\alpha\beta}+ E^{c}\wedge E^{\underline{\beta}} R_{\underline{\beta}\; c}{}^{\alpha\beta} +\tfrac{1}{2}
E^{c}\wedge E^{b} R_{bc}{}^{\alpha\beta}\; ,
\\
& & \nonumber \\ \label{bRff=N8c} R^{\dot{\alpha}\dot{\beta}} =
-\tfrac{1}{4}R^{ab}\tilde{\sigma}{}_{ab}^{\dot{\alpha}\dot{\beta}} & = & -\tfrac{1}{2}
{E}{}^{{\gamma}}_{i}\wedge {E}{}^{{\delta}}_{j} \epsilon_{{\gamma}{\delta}}\bar{M}{}^{\dot{\alpha}\dot{\beta}\,
ij} -\tfrac{1}{2} \bar{E}{}^{\dot{\beta}i} \wedge  \bar{E}{}^{\dot{\gamma}j} \left(\epsilon_{\gamma\delta}
\bar{N}^{\dot{\gamma}\dot{\delta}}_{ij} +2\delta_{(\dot\gamma}{}^{\dot\alpha}\delta_{\dot{\delta}
)}{}^{\dot\beta} \bar{S}_{ij}\right) \nonumber \\ & & \nonumber \\ & & +{E}{}^{{\gamma}}_{i}\wedge
\bar{E}{}^{\dot{\gamma}j} R^{i}_{{\gamma}\;\dot{\gamma}j }{}^{\dot{\alpha}\dot{\beta}} +E^{c}\wedge
E^{\underline{\beta}} R_{\underline{\beta}\; c}{}^{\dot{\alpha}\dot{\beta}} +\tfrac{1}{2} E^{c}\wedge E^{b}
R_{bc}{}^{\dot{\alpha}\dot{\beta}}\; .
\end{eqnarray}

Eqs.~(\ref{Fab:=}) and  (\ref{Fff=}) can be combined as
\begin{eqnarray}
\label{Fab=} F_{ab\; ij} & = & \tfrac{1}{2} \sigma_{ab}^{\alpha\beta}  F_{\alpha\beta\; ij} +\tfrac{1}{2}
\tilde{\sigma}{}_{ab}^{\dot{\alpha}\dot{\beta}} F_{\dot{\alpha}\dot{\beta}\; ij} = \tfrac{i}{4}
\sigma_{ab}^{\alpha\beta} M_{\alpha\beta\; ij} +\frac{ie^{i\beta}}{12\cdot 4!}  \varepsilon_{ij[3][3']}
\bar{\chi}^{[3]}\tilde{\sigma}{}_{ab}\bar{\chi}^{[3']} \; .
\end{eqnarray}


\end{document}